\documentclass[aps,prb,twocolumn,showpacs,floatfix,superscriptaddress,nofootinbib]{revtex4}
\usepackage{graphicx}
\usepackage{amssymb}
\usepackage{amsmath}
\usepackage{lmodern}
\usepackage{color}
\usepackage{hyperref}
\usepackage{empheq}
\usepackage{epstopdf}
\usepackage{amsmath}

\begin{document}

\title{The shocks in Josephson transmission line revisited}

\author{Eugene Kogan}
\email{Eugene.Kogan@biu.ac.il}
\affiliation{Department of Physics, Bar-Ilan University, Ramat-Gan 52900, Israel}
\affiliation{Donostia International Physics Center (DIPC)\\
San Sebastian/Donostia E-20018, Spain}

\begin{abstract}
We continue our previous studies of the shocks
in the   lossy Josephson transmission line (JTL).
The paper
consists of two parts. In the first part we  analyse the scattering of the "sound' (small amplitude small wave vector harmonic wave)   on the shock wave and calculate the reflection and the transmission coefficients.
In the second part we  show that the kinks, which we previously studied only in the lossless JTL, exist also in the lossy JTL and study
the similarities and the dissimilarities between  the shocks and the kinks there.
We  find that the nonlinear equation describing  the weak  kinks
and the weak shocks  can be integrated (in particular cases) in terms of elementary functions.
We also show that the profile of the shock in the lossy JTL demonstrates
oscillatory behavior with leading peaks resembling solitary waves if the losses are weak.
\end{abstract}

\date{\today}

\maketitle

\section{Introduction}

The interest in studies of  nonlinear electrical transmission lines, in particular of lossy nonlinear transmission lines, has started some time ago \cite{rosenau,chen,mohebbi}, but it became even more pronounced recently
\cite{ricketts,houwe,katayama,sekulic}. In ref.  \cite{malomed2}, one can find a very recent and complete review of studies of nonlinear electric transmission networks.

We studied previously the shock waves in the lossy Josephson transmission line (JTL) JTL \cite{kogan1,kogan2,kogan3} and kinks  and solitons in the lossless (actually, without any shunting at al) JTL \cite{kogan2,kogan3}.
The present work had several aims. First we would like to
analyse the interaction between the "sound" (small amplitude small wave vector harmonic wave) and the shock wave. Second  we would like to establish the relation between the shock waves and the kinks. And third, we would like to additionally study the weak  waves, and, in particular, to look for the cases when the nonlinear  equation, describing such waves in the JTL, can be integrated in terms of elementary functions.

 The rest of the article is constructed as follows.
In Section \ref{con} we rederive the circuit equations describing the JTL in the continuum approximation.
In  Section \ref{sou} we consider scattering of the "sound" wave
by the shock wave and calculate the appropriate reflection and transmission coefficients.
In Section \ref{unity} we show that the kinks which represent stationary solutions in the lossless JTL, can be also observed as weakly attenuating disturbances in the lossy JTL, which support stationary shock waves.  The connection between the shocks and the kinks in lossy JTL is revealed. We show that the profile of the shocks for the case of weak losses demonstrates soliton-like features.
We  also  integrate the wave equation  describing weak shocks (and kinks)
in terms of elementary functions for the specific value of the losses parameter.
We conclude in Section \ref{concl}. In the Appendix \ref{real} we present a physically appealing model of the JTL, composed of superconducting grains. In the Appendix \ref{dif} we formulate the condition for the applicability of the continuum approximation used in the paper. Some mathematical details are relegated to Appendices \ref{nomer} and \ref{ana}.

\section{The circuit equations: continuum approximation}
\label{con}

The discrete model of the  Josephson transmission line (JTL) is constructed from the identical Josephson junctions (JJs)  capacitors and resistors, as shown on Fig. \ref{trans5}.  (Possible physical realization of the model  is presented in the Appendix \ref{real}.)
We take
as the dynamical variables  the phase differences (which we for brevity will call just phases) $\varphi_n$ across the  JJs
and the voltages $v_n$ of the ground capacitors.
The circuit equations are
\begin{subequations}
\label{a10}
\begin{alignat}{4}
\frac{\hbar}{2e}\frac{d \varphi_n}{d t}&=v_{n-1}-v_{n} \label{a8a}\\
C\frac{dv_n}{dt} &=  I_c\sin\varphi_n- I_c\sin\varphi_{n+1}\nonumber\\
&+\left(\frac{1}{R_J}
+C_J\frac{d}{d t}\right)\frac{\hbar}{2e}\frac{d}{d t}\left(\varphi_{n}-\varphi_{n+1}\right),\label{a8b}
\end{alignat}
\end{subequations}
where    $C$ is the capacitance,  $I_c$ is the critical current of the JJ,
 and $C_J$ and $R_J$ are the capacitor and the ohmic resistor shunting the JJ.
\begin{figure}[h]
\includegraphics[width=\columnwidth]{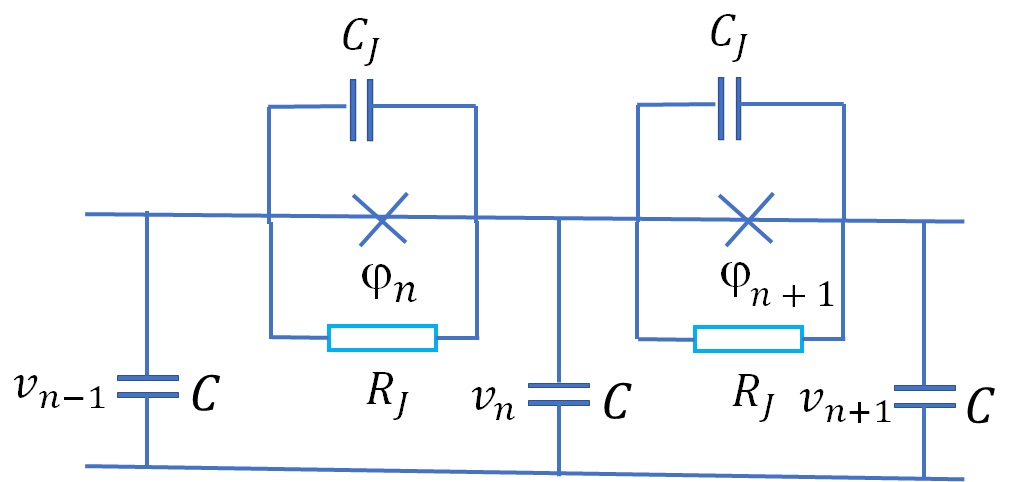}
%\vskip -.5cm
\caption{Discrete Josephson transmission line.}
 \label{trans5}
\end{figure}

In the continuum approximation we  treat $n$  as the continuous variable $Z$ and approximate the finite differences in the r.h.s. of the equations by the first derivatives with respect to $Z$,   after which the equations take the form
\begin{subequations}
\label{ve9c}
\begin{alignat}{4}
\frac{\partial \varphi}{\partial\tau}&= -\frac{\partial V}{\partial Z}, \label{vb0}\\
\frac{\partial V}{\partial\tau} &=  -\frac{\partial }{\partial Z}\left(\sin\varphi
+\frac{Z_J}{R_J}\frac{\partial\varphi}{\partial\tau}
+\frac{C_J}{C}\frac{\partial^2\varphi}{\partial\tau^2}\right).\label{vb}
\end{alignat}
\end{subequations}
where we  introduced the dimensionless time $\tau=t/\sqrt{L_JC}$ and the dimensionless voltage $V=v/(Z_JI_c)$; $L_J\equiv\hbar/(2eI_c)$ is the "inductance" of the  JJ and  $Z_J\equiv\sqrt{L_J/C}$ is the "characteristic impedance" of the JTL.
The condition for the applicability of the continuum approximation is formulated explicitly  in the Appendix \ref{dif}.

\section{The sound scattering by the shock wave}
\label{sou}

\subsection{The sound waves and the shock waves}

Because  (\ref{vb})  is nonlinear, the system   (\ref{ve9c}) has a lot of different types of solutions.
In this Section we'll be interested in only two types of those. First type -
small amplitude small wave vector harmonic waves
on a homogeneous background $ \varphi_0$.
For such waves Eq. (\ref{vb}) is simplified to
\begin{eqnarray}
\label{ve9bb}
\frac{\partial V}{\partial\tau} =  -\cos\varphi_0\frac{\partial \varphi}{\partial Z}.
\end{eqnarray}
We ignored the shunting terms in r.h.s. of (\ref{vb0}) because they contain higher order derivatives in comparison with the main term, and small wave vector means also small frequency.

The harmonic wave solutions of Eq. (\ref{ve9c}) (which, for brevity we'll call the sound)
are
\begin{subequations}
%\label{e9b}
\begin{alignat}{4}
\varphi&= \varphi_0+\varphi^{(h)} e^{ikz-i\omega \tau},\label{e9ba}\\
V&=V_0+ V^{(h)} e^{ikz-i\omega \tau},\label{e9bb}
\end{alignat}
\end{subequations}
where
\begin{eqnarray}
\label{uu}
\omega=\overline{u}\left(\varphi_0\right)k,\hskip 1cm \text{and}\hskip 1cm \overline{u}^2\left(\varphi_0\right)=\cos\varphi_0,
\end{eqnarray}
$\overline{u}\left(\varphi_0\right)$ being the normalized sound speed. In this paper the normalized  speed $\equiv$ physical speed times $\sqrt{L_J C}/\Lambda$,  where $\Lambda$ is the JTL period. Note that the stability of  a homogeneous background $ \varphi_0$  demands
\begin{eqnarray}
\label{k}
\cos\varphi_0>0.
\end{eqnarray}

The second type of solutions we'll be (mostly) interested in, is shock waves \cite{kogan1,kogan2}, which are (locally) the solutions satisfying the conditions
\begin{eqnarray}
\label{local}
\varphi(\tau,Z)=\varphi(\tau-Z/\overline{U}),\hskip .5cm
V(\tau,Z)=V(\tau-Z/\overline{U}).
\end{eqnarray}
 Substituting the ansatz (\ref{local}) into (\ref{ve9c})  and integrating  thus obtained equations with respect to $\tau$ from $-\infty$ to $+\infty$   we obtain
\begin{subequations}
\label{av78}
\begin{alignat}{4}
\overline{U}\left(\varphi_2-\varphi_1\right)&=V_2-V_1  ,\label{av78a}\\
\overline{U}\left(V_2-V_1\right)&= \sin\varphi_2-\sin\varphi_1,\label{av78b}
\end{alignat}
\end{subequations}
where $\varphi_1$ and $V_1$ are the phase and the voltage before the shock,  $\varphi_2$ and $V_2$ - after the shock, and
 $\overline{U}$ is the normalized shock wave speed.
Equation (\ref{av78}) is actually the Rankine-Hugoniot condition.
The obvious result of (\ref{av78}) is:
\begin{eqnarray}
\label{foa}
\overline{U}_{\varphi_2,\varphi_1}^2
=\frac{\sin\varphi_1-\sin\varphi_2}{\varphi_1-\varphi_2}.
\end{eqnarray}
Note that the shunting of the JJ doesn't influence the shock speed  \cite{kogan1,kogan2} but determines (as we will see explicitly in the next Section) the structure of the shock front.

In this Section we ignore the structure of the shock wave
and consider it as the discontinuities of the dynamical variables. Equations (\ref{av78}) connect these discontinuities   with the shock speed.

\subsection{The reflection and the transmission coefficients}

\begin{figure}[h]
\includegraphics[width=.8\columnwidth]{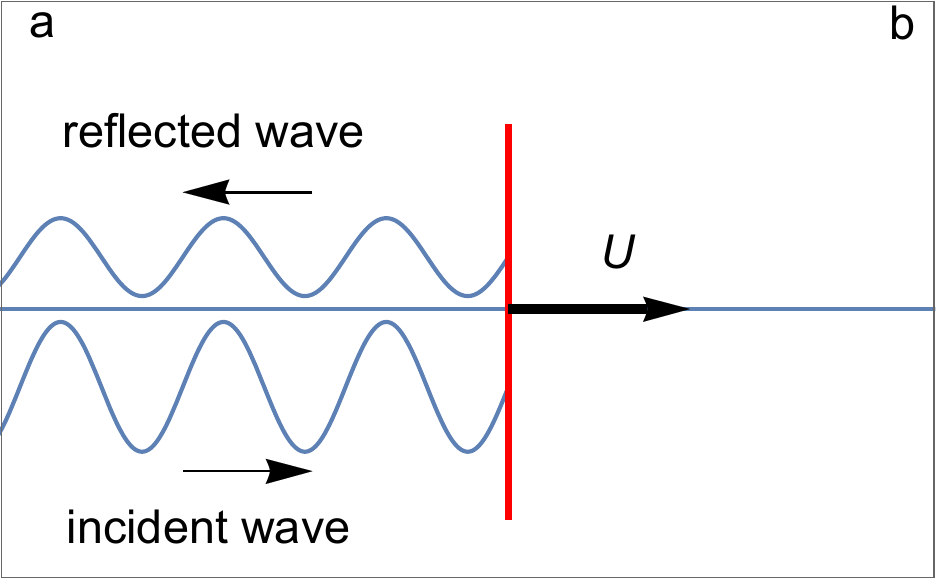}
%\vskip -.5cm
\caption{Reflection of a sound wave from a shock wave. The horizontal axis is the coordinate $Z$, the vertical axis - instantaneous value of the Josephson phase.}
\label{reflection}
\end{figure}

\begin{figure}[h]
\includegraphics[width=.8\columnwidth]{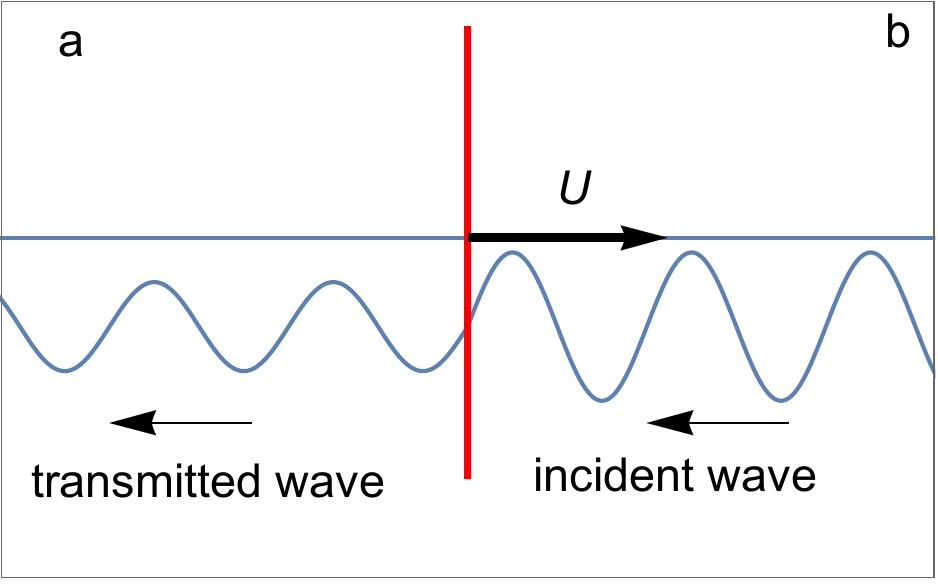}
%\vskip -.5cm
\caption{Transmission of a sound wave through a shock wave. The horizontal axis is the coordinate $Z$, the vertical axis - instantaneous value of the Josephson phase.}
\label{transmission}
\end{figure}

In this Section we'll be interested in two problems \cite{landau}.
The first one: A  sound wave is incident  from the rear  on a shock wave.
Determine the sound reflection coefficient $R$. The situation is shown in Fig. \ref{reflection}.
The second problem: A  sound wave is incident  from the front on a shock wave. Determine the sound
transmission coefficient $T$. The situation is shown in Fig. \ref{transmission}.
While formulating both problems we took into account the equation, which will be derived in Section \ref{unity}
\begin{eqnarray}
\overline{u}_b^2<\overline{U}_{\varphi_a,\varphi_b}^2<\overline{u}_a^2.
\end{eqnarray}
where
$\varphi_b$ and $\varphi_a$  are the phases before and after the shock in the absence of the sound respectively. Also,
\begin{eqnarray}
\label{ine}
\overline{U}_{\varphi_2,\varphi_1}&=\overline{U}_{\varphi_a,\varphi_b}
+\delta\overline{U}^{(r,t)}.
\end{eqnarray}

For the first problem mentioned above we have
\begin{subequations}
%\label{in}
\begin{alignat}{4}
\varphi_1&=\varphi_b,\label{ina}\\
V_1&=V_b,\\
\varphi_2&=\varphi_a+\varphi^{(in)}+\varphi^{(r)},\\
V_2&=V_a+V^{(in)}+ V^{(r)},\label{ind}
\end{alignat}
\end{subequations}
where   (in) stands for the incident sound wave and (r) for the reflected sound wave.
Substituting (\ref{ine}) - (\ref{ind})  into (\ref{av78a}), (\ref{av78b})
 in the first order approximation we obtain
\begin{subequations}
%\label{in2}
\begin{alignat}{4}
&\delta\overline{U}\left(\varphi_a-\varphi_b\right)
+\overline{U}\left(\varphi^{(in)}+\varphi^{(r)}\right)
= V^{(in)}+V^{(r)}  ,\\
&\delta\overline{U}\left(V_a-V_b\right)+\overline{U}\left( V^{(in)}+ V^{(r)}\right)
= \overline{u}^2(\varphi_a)\left(\varphi^{(in)}+\varphi^{(r)}\right).
\end{alignat}
\end{subequations}
Taking into account the relations
\begin{subequations}
\begin{alignat}{4}
V^{(in)}&=\overline{u}\left(\varphi_a\right) \varphi^{(in)},\\
 V^{(r)}&=-\overline{u}\left(\varphi_a\right) \varphi^{(r)}
\end{alignat}
\end{subequations}
(the difference in the signs is because of the opposite directions of propagation of the two waves) and excluding $\delta\overline{U}$ we obtain
\begin{eqnarray}
\label{rr}
R\equiv \frac{ \varphi^{(r)}}{ \varphi^{(in)}}=-\frac{\left[\overline{u}\left(\varphi_a\right)
-\overline{U}\right]^2}
{\left[\overline{u}\left(\varphi_a\right)+\overline{U}\right]^2}
=-\frac{\overline{u}_{in}^2}{\overline{u}_{r}^2},
\end{eqnarray}
where $\overline{u}_{in}=\overline{u}\left(\varphi_a\right)-\overline{U}$ is the speed of the incident sound wave relative to the shock wave, and $\overline{u}_{r}=\overline{u}\left(\varphi_a\right)+\overline{U}$ is the speed of the reflected sound wave relative to the shock wave.
As one could have expected, the modulus of the sound reflection coefficient is less than one, and it goes to zero when the intensity of the shock wave decreases, that is when $\varphi_a\to \varphi_b$,
in other words, when the shock wave itself nearly becomes the sound wave.

Now let us turn to the second problem.
We have
\begin{subequations}
%\label{out}
\begin{alignat}{4}
\varphi_1&=\varphi_b+\varphi^{(in)},\label{outa}\\
V_1&=V_b+ V^{(in)},\\
\varphi_2&=\varphi_a+\varphi^{(t)},\\
V_2&=V_a+ V^{(t)},\label{outd}
\end{alignat}
\end{subequations}
where (t) stands for the transmitted wave.
Substituting (\ref{ine}), (\ref{outa}) - (\ref{outd}) into (\ref{av78a}), (\ref{av78b}),
in the first order approximation we obtain
\begin{subequations}
%\label{in2}
\begin{alignat}{4}
\delta\overline{U}\left(\varphi_a-\varphi_b\right)
+\overline{U}\left(\varphi^{(t)}-\varphi^{(in)}\right)
= V^{(t)}- V^{(in)} \\
\delta\overline{U}\left(V_a-V_b\right)+\overline{U}\left( V^{(t)}- V^{(in)}\right)
\nonumber\\
= \overline{u}^2(\varphi_a) \varphi^{(t)}- \overline{u}^2(\varphi_b) \varphi^{(in)}.
\end{alignat}
\end{subequations}
Taking into account the relations
\begin{subequations}
\begin{alignat}{4}
V^{(in)}&=-\overline{u}\left(\varphi_b\right) \varphi^{(in)},\\
V^{(t)}&=-\overline{u}\left(\varphi_a\right) \varphi^{(t)}
\end{alignat}
\end{subequations}
 and excluding $\delta\overline{U}$ we obtain
\begin{eqnarray}
\label{rr4}
T\equiv \frac{ \varphi^{(t)}}{ \varphi^{(in)}}
=\frac{\left[\overline{u}\left(\varphi_b\right)+\overline{U}\right]^2}
{\left[\overline{u}\left(\varphi_a\right)+\overline{U}\right]^2}
=\frac{\overline{u}_{in}^2}{\overline{u}_{t}^2},
\end{eqnarray}
where $\overline{u}_{in}=\overline{u}\left(\varphi_b\right)
+\overline{U}$ is the speed of the incident sound wave relative to the shock wave, and $\overline{u}_{t}=\overline{u}\left(\varphi_a\right)+\overline{U}$ is the speed of the transmitted sound wave relative to the shock wave.
As one could have expected, the sound transmission coefficient is less than one, and goes to one when the intensity of the shock wave decreases, that is when $\varphi_a\to \varphi_b$.

Looking back at the derivation of (\ref{rr}) and (\ref{rr4}) we understand that the equations will be valid also for a generalized Josephson law for the supercurrent $I_s$:
\begin{eqnarray}
\label{general}
I_s=I_cf(\varphi).
\end{eqnarray}
where $f$ is a (nearly) arbitrary function. The difference from the case considered above is that the sound speed in the general  case is
\begin{eqnarray}
\label{general2}
\overline{u}^2\left(\varphi_0\right)=f'(\varphi_0),
\end{eqnarray}
and the shock speed is given by the equation
\begin{eqnarray}
\label{foa2}
\overline{U}_{\varphi_2,\varphi_1}^2
=\frac{f(\varphi_1)-f(\varphi_2)}{\varphi_1-\varphi_2}.
\end{eqnarray}

\section{The  shocks and the kinks}
\label{unity}

\subsection{The  travelling waves}

In this Section we would like to study the structure of the shock wave, so we return to  Eq. (\ref{ve9c}).
Consider a solution  which for $\tau\in (-\infty,+\infty)$ stays in the finite region of the phase space. The limit cycles are excluded for our problem,
and strange attractors are excluded  in a 2d phase space in general \cite{strogatz}. Hence the trajectory begins  in a fixed point and ends  in a fixed point
\begin{eqnarray}
 \label{12}
\lim_{\tau\to -\infty}\varphi&=\varphi_1,  \hskip 1cm
\lim_{\tau\to +\infty}\varphi=\varphi_2.
\end{eqnarray}

In this Section we will use  the  ansatz (\ref{local})  globally, i.e. as describing  the travelling wave, $\overline{U}$ being its  speed.
Using the ansatz we can write down Eqs. (\ref{ve9c}) as
\begin{subequations}
\label{vc}
\begin{alignat}{4}
\overline{U}\frac{d \varphi}{d\tau}&= \frac{d V}{d \tau}, \label{vca}\\
\overline{U}\frac{d V}{d\tau} &=  \frac{d }{d \tau}\left(\sin\varphi
+\frac{Z_J}{R_J}\frac{d\varphi}{d\tau}
+\frac{C_J}{C}\frac{d^2\varphi}{d\tau^2}\right).\label{vcb}
\end{alignat}
\end{subequations}
Note that from Eq. (\ref{vca})  follows that in the travelling wave the voltage $V$ is connected to the Josephson phase $\varphi$ in a very simple way
\begin{eqnarray}
\label{vca4}
V=\overline{U} \varphi
\end{eqnarray}
(of course, an arbitrary constant can be added to the r.h.s. of (\ref{vca4})).

Excluding $V$ from (\ref{vc}) we obtain
\begin{eqnarray}
\label{c}
\overline{U}^2\frac{d \varphi}{d\tau}= \frac{d }{d \tau}\left(\sin\varphi
+\frac{Z_J}{R_J}\frac{d\varphi}{d\tau}
+\frac{C_J}{C}\frac{d^2\varphi}{d\tau^2}\right).
\end{eqnarray}
Integrating (\ref{c}) we get
\begin{eqnarray}
\label{system26}
\frac{d^2\varphi}{d \tilde{\tau}^2 }+\gamma\frac{d\varphi}{d\tilde{\tau}}
+\sin\varphi=\overline{U}^2\varphi+F,
\end{eqnarray}
where
$\tilde{\tau}\equiv \tau\sqrt{C/C_J}=t/\sqrt{L_JC_J}$,
$\gamma\equiv\sqrt{L_J/C_J}\big/R_J$  is the damping coefficient, and $F$  is
the constant of integration.
Equation (\ref{system26}) reminds the equation
\begin{eqnarray}
\label{kk}
\frac{d^2\varphi}{d \tilde{\tau}^2 }+\gamma\frac{d\varphi}{d\tilde{\tau}}
+\sin\varphi=I/I_c,
\end{eqnarray}
describing current biased JJ within the RCSJ model \cite{tinkham}.

\subsection{ The kinks vs. the shocks}
\label{uni}

Taking into account the boundary conditions   (\ref{12})
and the equation for the shock speed (\ref{foa}),
we obtain that $F$ in Eq. (\ref{system26}) is
\begin{eqnarray}
\label{system3}
F=\frac{\varphi_1\sin\varphi_2-\varphi_2\sin\varphi_1}
{\varphi_1-\varphi_2}.
\end{eqnarray}
 Hence Eq.  (\ref{system26}) can be written down as
\begin{eqnarray}
\label{system4}
\frac{d^2\varphi}{d \tilde{\tau}^2 }+\gamma\frac{d\varphi}{d\tilde{\tau}}
=-\frac{d\Pi(\varphi)}{d\varphi},
\end{eqnarray}
where
\begin{eqnarray}
\label{w}
\Pi(\varphi)=\frac{\left(\varphi-\varphi_1\right)^2\sin\varphi_2-
\left(\varphi-\varphi_2\right)^2\sin\varphi_1}{2(\varphi_1-\varphi_2)}
-\cos\varphi.
\end{eqnarray}
Equation (\ref{system4}) is Newton equation,  describing  motion with friction of the fictitious particle in the potential well $\Pi(\varphi)$. The motion starts at $\tilde{\tau}=-\infty$ at $\varphi=\varphi_1$ and ends at $\tilde{\tau}=+\infty$ at $\varphi=\varphi_2$, and
\begin{eqnarray}
\label{ineq}
\Pi(\varphi_2)<\Pi(\varphi_1).
\end{eqnarray}

\begin{figure}[h]
\includegraphics[width=.8\columnwidth]{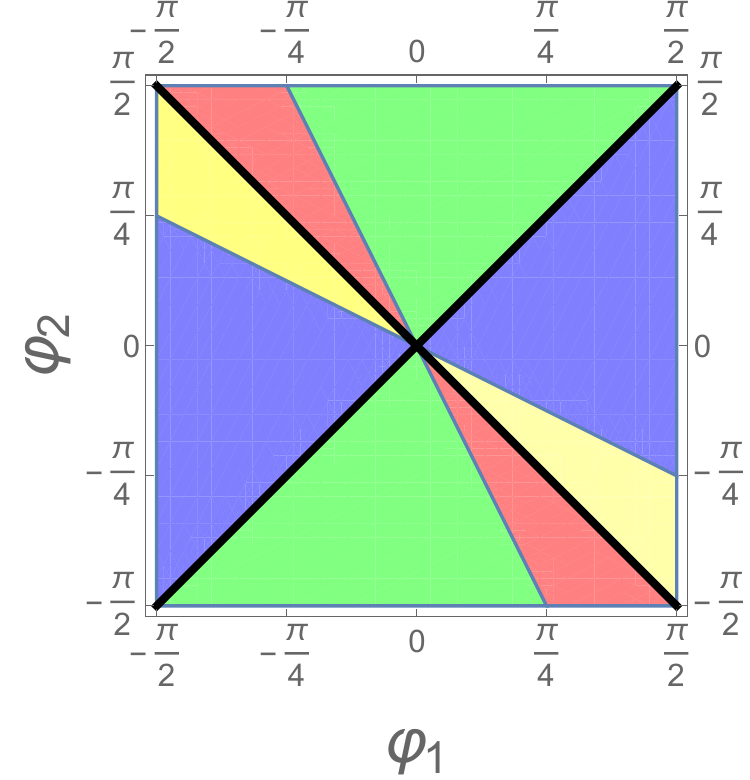}
\caption{The phase plane of the boundary conditions $(\varphi_1,\varphi_2)$. Blue regions correspond to the shock wave moving to the right,
green regions - to the left.
Yellow regions correspond to the kink moving to the right,   red regions - to the left.
The thick black line $\varphi_2=-\varphi_1$ corresponds to the kink, the thick black line $\varphi_2=\varphi_1$ - to the soliton which can exist only in the bare-bones JTL and  propagate in both directions. }
 \label{col}
\end{figure}

Because of the   invariance of the system when all phases are shifted by $2\pi$,
it is enough to consider $\varphi_1\in (-\pi,\pi)$. Because of the condition (\ref{k}), we should consider only $\varphi_1\in (-\pi/2,\pi/2)$, and because the physics is obviously symmetric with respect to simultaneous inversion of all phases $\varphi \to -\varphi$, in the following
 we  consider explicitly (everywhere apart from Fig. \ref{col}) only
$\varphi_1\in (0,\pi/2)$.

We consider in this paper only the case $\varphi_2\in (-\pi/2,\pi/2)$.
Because the sine function monotonically increases between $-\pi/2$ and $\pi/2$, the r.h.s. of (\ref{foa}) is always positive in this case.
From (\ref{ineq})  follows $-\varphi_1<\varphi_2<\varphi_1$.

\begin{figure}[h]
\includegraphics[width=.6\columnwidth]{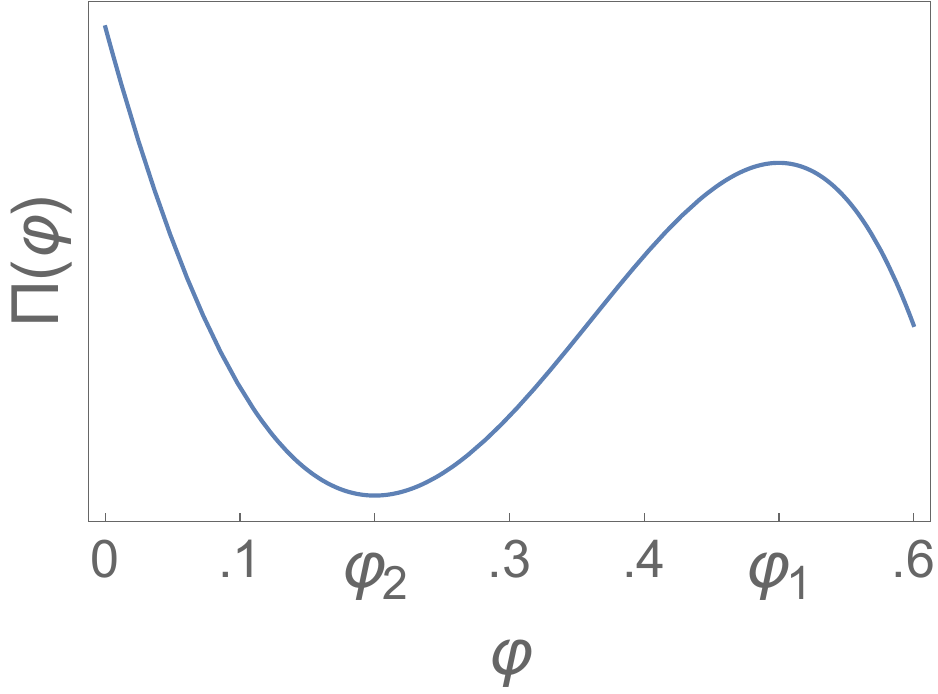}
\caption{The potential  $\Pi(\varphi)$, as given by (\ref{w}), for $\varphi_1=.5$, $\varphi_2=.2$. }
 \label{we2}
\end{figure}

If   $\varphi_2$  is  positive, it is inevitably the point of a minimum of the potential.
In fact, the stationary points of the potential are given by the equation
\begin{eqnarray}
\label{rhs}
\sin\varphi=\overline{U}^2\varphi+F.
\end{eqnarray}
Because $\sin\varphi$ is concave downward
for $0 < \varphi< \pi/2$, the straight line, crossing the sine curve at the points $\pi/2>\varphi_1,\varphi_2>0$ can't cross the curve in  between. Hence there are no stationary points between $\varphi_1$ and $\varphi_2$.
The potential  $\Pi(\varphi)$  for positive $\varphi_2$  is illustrated in Fig. \ref{we2}.

\begin{figure}[h]
\includegraphics[width=.45\columnwidth]{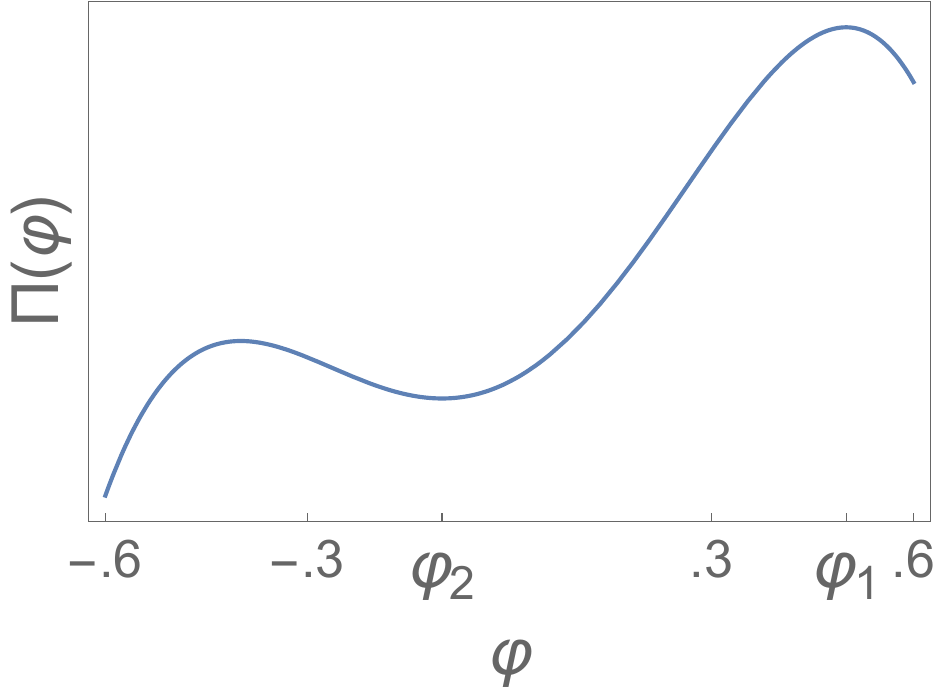}
\includegraphics[width=.45\columnwidth]{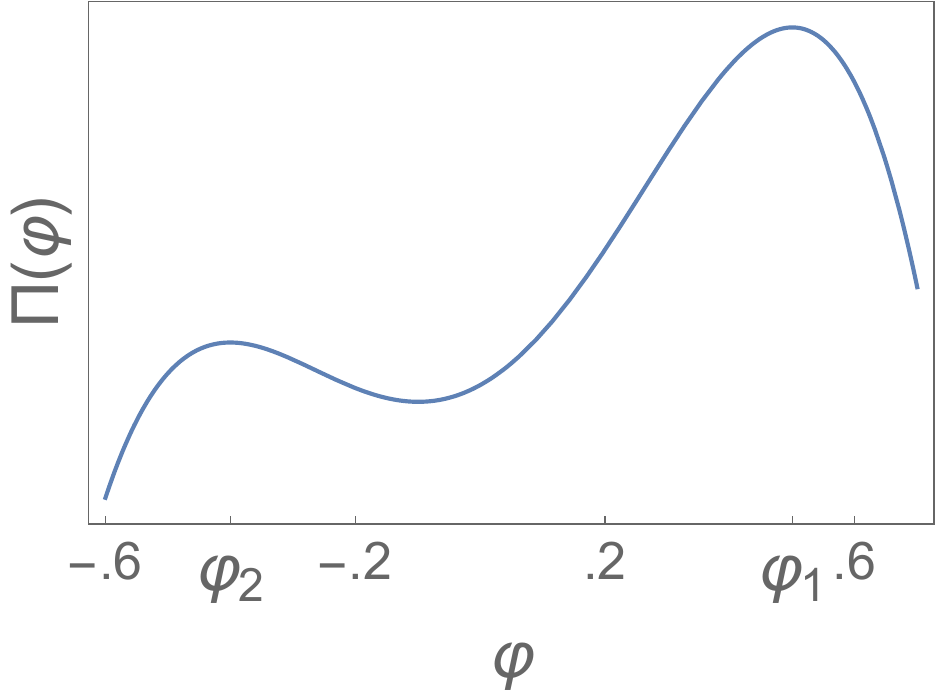}
\caption{The potential  $\Pi(\varphi)$, as given by (\ref{w}), for $\varphi_1=.5$, $\varphi_2=-.2$ (left) and for $\varphi_1=.5$, $\varphi_2=-.4$ (right). In the former case $\varphi_2$ corresponds to the minimum of the potential, in the latter - to the maximum.}
 \label{we}
\end{figure}

On the other hand, for  $\varphi_2<0$  the potential  $\Pi(\varphi)$ can have either a minimum or a maximum at $\varphi_2$, as it  is
 illustrated in  Fig. \ref{we}.
Looking at  Fig.  \ref{we} (left)  we realize that for  the solution with $\varphi_1$ and $\varphi_2$ having opposite signs to exist, the effective friction coefficient $\gamma$ should be large enough to prevent escape of the particle above the potential barrier to the left of $\varphi_2$.
(There is no such restriction for the shock wave with $\varphi_1$ and $\varphi_2$ having the same sign, because in this case the left potential barrier is higher than the right one, as it is illustrated in Fig. \ref{we2}.)

\begin{figure}[h]
\includegraphics[width=.8\columnwidth]{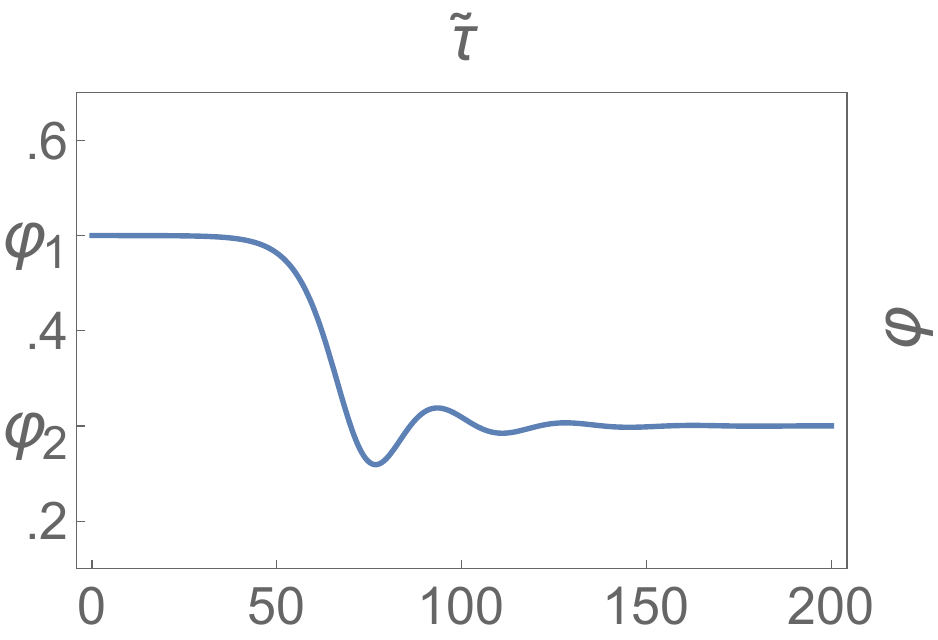}
\caption{Numerical solution of (\ref{system4}) for   $\varphi_1=.5$, $\varphi_2=.3$
and $\gamma=.1$. }
 \label{sho}
\end{figure}

The minimum of the potential at $\varphi_2$ situation
corresponds to the shock wave  and was discussed at length in our previous publications \cite{kogan1,kogan2}. In Appendix \ref{nomer} we explain how
Eq.  (\ref{system26})  in this case can  be  integrated numerically. The result of such integration is presented in Fig. \ref{sho}.

We considered previously  the potential
maximum at $\varphi_2$ case only   in the absence of shunting \cite{kogan2}.
We called such travelling waves the kinks.
Now we understand that similar kinks exist also in the lossy JTL (for $-\varphi_1<\varphi_2<\varphi_1/2$).
Looking at  Fig.  \ref{we} (right), presenting the potential for the kink,  we realize, that  since the particle  stops at the unstable equilibrium point, for  the  kink to exist,
 the fine tuning   is necessary - the parameters $\varphi_1$, $\varphi_2$ and $\gamma$ should satisfy definite relation. For weak kinks such relation will be obtained in section \ref{weak}.
Thus in the absence of losses ($\gamma=0$),  only the kinks with $\varphi_2=-\varphi_1$, are possible \cite{kogan2}.

We can find the boundary between the two cases  considered above (when $\varphi_2$ is an inflexion point of the potential)  by equating
the second derivative  of the potential at the point $\varphi_2$ to zero
\begin{eqnarray}
\label{in}
\overline{U}^2-\overline{u}^2\left(\varphi_2\right)=0.
\end{eqnarray}
The approximate solution of (\ref{in}) is $\varphi_2=-\varphi_1/2$.

Everywhere above we considered the travelling wave going to the right, but, of course, by interchanging $\varphi_1$ and $\varphi_2$  (also in the inequality (\ref{ineq})) we obtain the wave going to the left. So the conditions for the shocks and for the kinks
in the whole phase plane of the boundary conditions $(\varphi_1,\varphi_2)$
are shown in Fig. \ref{col}. Two additional straight lines on this Figure $\varphi_2=-\varphi_1$ and $\varphi_2=\varphi_1$ present the kinks and the solitons respectively, which can exist in the bare-bones (unshunted) JTL \cite{kogan2} and propagate in both directions.

Following the venerable tradition to tell the same story twice, we will now present an  approach to integration of the equation (\ref{c}), alternative to that we used above.
Introducing  the voltage at the Josephson junction
\begin{eqnarray}
\label{def}
E= \frac{d\varphi}{d\tilde{\tau}}
\end{eqnarray}
as the new dependent variable,
considering $\varphi$ as the independent variable
and dividing both parts by the common multiplier $E$,
we reduce the order of  Eq. (\ref{c})  and write it down as
\begin{eqnarray}
\label{22}
\frac{1}{2}\frac{d^2E^2}{d\varphi^2}+\gamma\frac{dE}{d\varphi}=
\overline{U}^2-\overline{u}^2(\varphi).
\end{eqnarray}
This equation will be used in Section \ref{weak} to study weak kinks(shocks).

Note that in the framework of the quasi-continuum
approximation E is (up to a numerical multiplier) just
the voltage on the JJ. Thus  Eq.
(36) gives the voltage on the JJ as
the function of the Josephson phase.

\subsection{The shocks speed vs. the kinks speed}

For the shock   $\varphi_1$ is  the point of a maximum of $\Pi(\varphi)$
and $\varphi_2$ is  the point of a minimum.
Hence
the second derivative of the potential with respect to $\varphi$ is  negative
at $\varphi_1$ and positive at $\varphi_2$.   Taking into account  Eq. (\ref{foa}), we obtain
\begin{eqnarray}
\label{30}
\overline{u}^2(\varphi_2)>\overline{U}^2_{\varphi_2,\varphi_1}>\overline{u}^2(\varphi_1).
\end{eqnarray}
The inequalities (\ref{30}) reflect the well-known in the nonlinear
waves theory fact: the shock speed is smaller than the sound
speed in the region behind the shock but larger than the sound
speed in the region before the shock \cite{whitham}.

From the inequalities (\ref{30}) follows that a shock  can not split into two shocks moving in the same direction.  Actually we can make even stronger statement: two shocks moving in the same direction will merge. In fact,  let there is the first shock   $\varphi_0\leftarrow\varphi_1$ and the second shock $\varphi_2\leftarrow\varphi_0$  behind it. Because of inequalities (\ref{30}) the speed of the second shock is larger,  and the speed of the first shock is smaller than $\overline{u}(\varphi_0)$.  The statement is proved.
Note that due to a one-dimensional nature of our problem we don't have to consider the corrugation instability of the shock wave
\cite{dyakov1,lubchich,semenko,robinet,bates}.

On the other hand,  two shocks
going in the opposite directions may well coexist. From that we understand that any given boundary conditions $\varphi_1,\varphi_2$
are compatible both with a single shock (kink) $\varphi_2\leftarrow\varphi_1$ and with two shocks
$\varphi_0\leftarrow\varphi_1$ and $\varphi_2\rightarrow \varphi_0$ (with any
$\varphi_0$, satisfying the conditions $|\varphi_3|<|\varphi_1|,|\varphi_2|$)
going in the opposite directions.  The solution with $\varphi_0=0$  looks especially appealing.
This dichotomy   will (hopefully) be analyzed in the next paper.

For the kink both  $\varphi_1$
and $\varphi_2$ are  the points of minima.
Hence
the second derivative of the potential with respect to $\varphi$ is  positive at both points.  Thus
\begin{eqnarray}
\label{31}
\overline{U}^2_{\varphi_2,\varphi_1}>\overline{u}^2(\varphi_2)>\overline{u}^2(\varphi_1).
\end{eqnarray}
The kink is supersonic  from the point of view both of the region before and after it.

\subsection{ The quasi-solitons within the shocks}

We studied  previously the kinks and the solitons in the absence of  damping \cite{kogan2}.
In Section \ref{uni} we have shown that the kinks  exist also in the presence of   damping.  So what about the solitons?

Looking at the Fig. \ref{we2} we understand that in the absence of  damping ($R_J=\infty\rightarrow\gamma=0$) the motion of the fictitious particle starting and ending at $\varphi=\varphi_1$  is possible. This corresponds to the boundary conditions (\ref{12}) with $\varphi_2=\varphi_1$ (the soliton).
When we switch on the  damping the solitons don't exist
any more, because the particle can't return to its initial position. However,  for weak damping, the particle after leaving the initial equilibrium position $\varphi_1$ will nearly return to this position after reflection from the opposite wall of the potential well
(see Fig. \ref{we2}).
The part of the shock described by that motion will look very much like the soliton.
 The point of the reflection $\varphi_0$ can be found from the equation
\begin{eqnarray}
\label{pio}
\Pi(\sin\varphi_0) \approx \Pi(\sin\varphi_1).
\end{eqnarray}
Substituting the formula for the potential energy
(\ref{w}) into (\ref{pio})  we can connect between $\varphi_2$ and  $\varphi_0$,
thus obtaining   an alternative to  (\ref{foa}) expression for the wave speed
\begin{eqnarray}
\label{solist}
\overline{U}^2(\varphi_1,\varphi_0)
=2\frac{\cos\varphi_1-\cos\varphi_0+(\varphi_1-\varphi_0)\sin\varphi_1}
{(\varphi_1-\varphi_0)^2}.
\end{eqnarray}
Note that the leading peak will be followed by the other ones, only (because of the continuous decrease of energy of the particle) the distance between the successive peaks will be decreasing and each successive  peak will look less and less like the real soliton.

\subsection{Weak  kinks and elementary  weak shocks}
\label{weak}

For weak  wave, characterized by the condition
$\varphi_1-\varphi_2\ll 1$,  the r.h.s. of (\ref{system4})
can be approximated as
\begin{eqnarray}
\label{ystem44}
-\frac{d\Pi(\varphi)}{d\varphi}
=\alpha(\varphi-\varphi_1)(\varphi-\varphi_2)
(\varphi+\varphi_3),
\end{eqnarray}
where
\begin{eqnarray}
\overline{\varphi}\equiv \frac{\varphi_1+\varphi_2}{2},\hskip .5cm
\varphi_3\equiv 3\tan\overline{\varphi}-\overline{\varphi},\hskip .5cm \alpha\equiv \frac{\cos\overline{\varphi}}{6},
\end{eqnarray}
and  (\ref{system4}) can be simplified to the damped Helmholtz-Duffing (dHD) equation
\begin{eqnarray}
\label{system444}
\varphi_{\tilde{\tau}\tilde{\tau}}+\gamma\varphi_{\tilde{\tau}}
=\alpha(\varphi-\varphi_1)(\varphi-\varphi_2)(\varphi+\varphi_3).
\end{eqnarray}
For  $\varphi_2<-\varphi_3$,  Eq. (\ref{system444})
describes the kink, for $-\varphi_3<\varphi_2$
- the shock.

Using the results of  Appendix \ref{ana} we state that
if the constants in (\ref{system444}) satisfy the condition
\begin{eqnarray}
\label{g}
\gamma=\sqrt{\frac{\alpha}{2}}(\varphi_1+\varphi_2+2\varphi_3)
=\sqrt{3\cos\overline{\varphi}}\tan\overline{\varphi},
\end{eqnarray}
the solution of (\ref{system444})  satisfying the   boundary conditions
(\ref{12}) is
\begin{eqnarray}
\label{fi}
\varphi(\tilde{\tau})=\overline{\varphi}+\frac{\Delta\varphi}{2}
\tanh\left(\beta\tilde{\tau}\right),
\end{eqnarray}
where $\Delta\varphi=\varphi_1-\varphi_2$ and
\begin{eqnarray}
\label{g60}
\beta=\sqrt{\frac{\alpha}{8}}\Delta\varphi
=\sqrt{\frac{\cos\overline{\varphi}}{48}}\Delta\varphi.
\end{eqnarray}

Note that for $\varphi_2<-\varphi_3$, Eq. (\ref{g}) is both the condition of the kink existence (fine tuning we talked about previously) and of the kink being given by the elementary function. The shock doesn't demand find tuning, so  for $-\varphi_3<\varphi_2$,  Eq. (\ref{g}) is only the condition of the shock being given by the elementary function.

If the assumption $\Delta\varphi\ll 1$ is strengthened  to $\Delta\varphi\ll\tan\overline{\varphi}$,  Eq. (\ref{system444}) can be approximated as
\begin{eqnarray}
\label{sm}
\varphi_{\tilde{\tau}\tilde{\tau}}+\gamma\varphi_{\tilde{\tau}}
=\frac{\sin\overline{\varphi}}{2}(\varphi-\varphi_1)(\varphi-\varphi_2).
\end{eqnarray}
It is shown in  Appendix \ref{ana}
the  solution of  (\ref{sm})  satisfying the   boundary conditions (\ref{12}) is
\begin{eqnarray}
\label{fi8}
\varphi(\tilde{\tau})=\varphi_2+\frac{\Delta\varphi}
{\left[\exp(\beta'\tilde{\tau})+1\right]^2},
\end{eqnarray}
where
\begin{eqnarray}
\label{eb}
\beta'=\sqrt{\Delta\varphi\sin\overline{\varphi}/12},\hskip 1cm \gamma=5\beta'.
\end{eqnarray}
In the vicinity of $\varphi=0$ and for $\Delta\varphi\ll\tan\overline{\varphi}$, the condition  (\ref{g})
corresponds to strongly overdamped oscillator, and
the condition (\ref{eb}), to slightly overdamped oscillator.
The shock described by Eq. (\ref{sm})  exists for any value of $\gamma$. Equation (\ref{eb}) is only the condition of it being given by the elementary function.

In both particular cases studied above, because of the overdamped nature of the  oscillations, the shock wave is monotonic.
Note that the detailed study of the shock wave structure  in a strongly nonlinear lattice with viscous dissipation was presented in Ref. \cite{nesterenko} where, in particular, the dichotomy between the oscillatory and the monotonic shock waves was analysed quantitatively. We, however, postpone a similar analysis for the shocks in the JTL until the next publication.

It is interesting to see how the results obtained above can be recovered in the framework of the approach based on Eq. (\ref{22}).
If $\varphi_1$ and $\varphi_2$ are close enough to each other, we  can approximate $\cos\varphi$ between $\varphi_1$ and $\varphi_2$   as a second degree polynomial in $\varphi$
\begin{eqnarray}
\label{delo}
\cos\varphi=\cos\overline{\varphi}-\varphi'\sin\overline{\varphi}
-\frac{\varphi'^2}{2}\cos\overline{\varphi},
\end{eqnarray}
where $\varphi'=\varphi-\overline{\varphi}$.
The boundary conditions (\ref{12}) obviously give
\begin{eqnarray}
\label{vk7}
E\left(\varphi_{1,2}\right)=0,
\end{eqnarray}
thus we may try the solution of (\ref{22}) in the form
\begin{eqnarray}
\label{m}
E(\varphi)=\psi \left(\varphi-\varphi_1\right)
\left(\varphi-\varphi_2\right),
\end{eqnarray}
where $\psi$ is a constant.
Substituting (\ref{m})  into (\ref{22})  we  obtain that (\ref{m}) is indeed the solution, provided Eq.  (\ref{g}) is valid and $\psi^2=\cos\overline{\varphi}/12$.
To find $\varphi(\tilde{\tau})$ we have to solve equation
\begin{eqnarray}
\label{555}
\frac{d\varphi}{d\tilde{\tau}}=\sqrt{\frac{\cos\overline{\varphi}}{12}}
\left(\varphi-\varphi_1\right)
\left(\varphi-\varphi_2\right).
\end{eqnarray}
The solution   is Eq. (\ref{fi}).

When $\Delta\varphi\ll \overline{\varphi}$ we can keep in the r.h.s. of (\ref{delo}) only the first two terms and present (\ref{22}) (with the accepted precision) as
\begin{eqnarray}
\label{226}
\frac{1}{2}\frac{d^2E^2}{d\varphi^2}+\gamma\frac{dE}{d\varphi}=
\overline{U}^2-\cos\overline{\varphi}
+\left(\varphi-\varphi_2\right)\sin\overline{\varphi}.
\end{eqnarray}
In this case the  solution of (\ref{226})  satisfying the boundary conditions Eq. (\ref{vk7})  is
\begin{eqnarray}
\label{an}
E=\chi \left(\varphi-\varphi_2\right)\left[\left(\varphi-\varphi_2\right)^{1/2}
-\left(\Delta\varphi\right)^{1/2}\right].
\end{eqnarray}
Substituting (\ref{an})  into (\ref{226}) and equating coefficients before the same powers of $\left(\varphi-\varphi_2\right)^{1/2}$ in the l.h.s and in the r.h.s. of the equation, we  obtain (\ref{eb}) and
$\chi^2=\sin\overline{\varphi}/3$.
The function $E(\varphi)$ being found, we can find the function $\varphi(\tilde{\tau})$ by solving equation
\begin{eqnarray}
\frac{d\varphi}{d\tilde{\tau}}=\sqrt{\frac{\sin\overline{\varphi}}{3}} \left(\varphi-\varphi_2\right)\left[\left(\varphi-\varphi_2\right)^{1/2}
-\left(\Delta\varphi\right)^{1/2}\right].
\end{eqnarray}
The solution is given by (\ref{fi8}).

\section{Conclusions}
\label{concl}

We considered the interaction of the sound waves with the shock waves in the JTL.
The formulas for the reflection  and the transmission coefficients turned out to be very simple and appealing.

We established the relation between the shocks existing in the lossy JTL and the kinks which, as we now understand, exist both in the lossy and in the lossless JTL.
We also have shown that the profile of the shock in the lossy JTL demonstrates
oscillatory behavior with leading peaks resembling solitary waves if the losses are weak.

We  found that the nonlinear equation describing  the weak  kinks
and the weak shocks  can be integrated (in particular cases) in terms of elementary functions.

\begin{appendix}

\section{The JTL composed of superconducting grains}
\label{real}

\begin{figure}[h]
\includegraphics[width=\columnwidth]{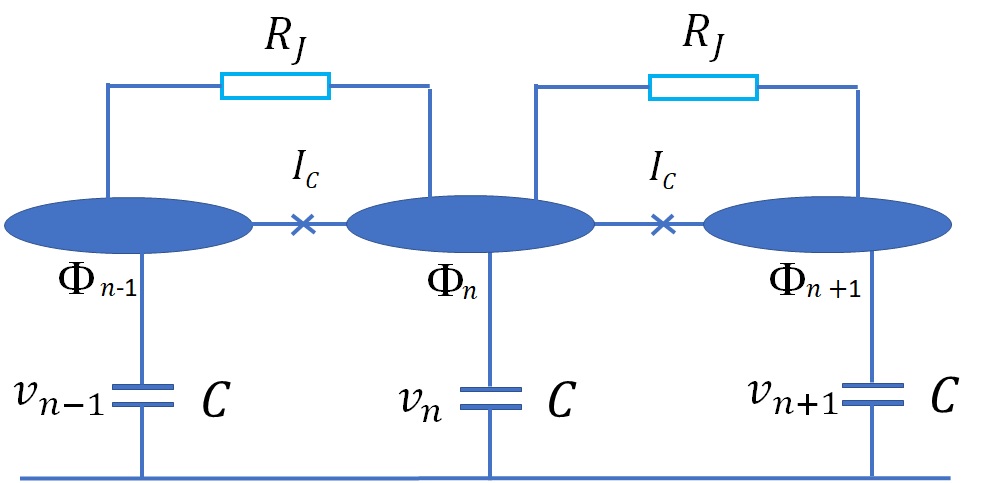}
\caption{Josephson transmission line  composed of superconducting grains}
 \label{gr1}
\end{figure}

A physically appealing model of the JTL composed of superconducting grains is presented in Fig. \ref{gr1} (for simplicity  we ignored the shunting capacitor).
Here, we take
as the dynamical variables  the phases at the grains $\Phi_n$ and the potentials of the grains $V_n$. The circuit equations are
\begin{subequations}
\label{ne}
\begin{alignat}{4}
\frac{\hbar}{2e}\frac{d \Phi_n}{d t}&=v_n \label{nea}\\
C\frac{dv_n}{dt} &=  I_c\sin\left(\Phi_{n-1}- \Phi_{n}\right)
-I_c\sin\left(\Phi_{n}- \Phi_{n+1}\right)\nonumber\\
&+\frac{1}{R_J}\left(v_{n-1}-2v_{n}+v_{n+1}\right).\label{neb}
\end{alignat}
\end{subequations}
We realise that Eq. (\ref{a10}) follows from Eq. (\ref{ne})
if we substitute $\varphi_n=\Phi_{n-1}-\Phi_n$.
Also, if we exclude $v_n$ from (\ref{nea}), (\ref{neb}) we obtain
\begin{eqnarray}
\label{mal}
\frac{d^2 \Phi_n}{d \tau^2}&=&  \sin\left(\Phi_{n-1}- \Phi_{n}\right)
-\sin\left(\Phi_{n}- \Phi_{n+1}\right)\nonumber\\
&+&\frac{Z_J}{R_J}\frac{d}{d\tau}\left(\Phi_{n-1}-2\Phi_{n}+\Phi_{n+1}\right),
\end{eqnarray}
which is the particular case of the Fermi-Pasta-Ulam-Tsingou  equation (with losses).

It is interesting to compare (\ref{mal}) with the equation from Ref. \cite{malomed}, describing the chain of interacting particles with friction
\begin{eqnarray}
m\frac{d^2y_n}{d\tau^2}
=&-&\frac{\partial}{\partial y_n}\left[U\left(y_{n-1}-y_{n}\right)
+U\left(y_{n+1}-y_{n}\right)\right]  \nonumber\\
&-&\gamma\frac{dy_n}{d\tau},
\end{eqnarray}
where  $y_n$ are  displacements of particles in the chain and $U(z)$ is the potential of the interparticle interaction.
Comparison shows  different character of the losses in the systems.

It is also interesting to compare the  one-dimensional Josephson-junction array,  described by
  the discretized version of the perturbed sine-Gordon
equation \cite{ustinov}
\begin{eqnarray}
\label{us}
\frac{d^2\varphi_n}{d\tau^2}+ \gamma\frac{d\varphi_n}{d\tau}+\sin\varphi_n
&-&\frac{1}{ a^2}\left(\varphi_{n-1}
-2\varphi_n+\varphi_{n+1}\right)\nonumber\\
&=&I/I_c,
\end{eqnarray}
with the equation obtained by excluding $v_{n}$ from  (\ref{a10})
\begin{eqnarray}
\label{di}
\frac{d ^2\varphi_n}{d \tau^2}=
\sin\varphi_{n+1}-2\sin\varphi_n+\sin\varphi_{n+1}\nonumber\\
+\left(\frac{Z_J}{R_J}
+\frac{C_J}{C}\frac{d}{d\tau}\right)\frac{d}{d\tau}
\left(\varphi_{n+1}-2\varphi_{n}+\varphi_{n+1}\right) \label{b}
\end{eqnarray}
shows  that the nature of nonlinearity in the systems is different.
Neither does (\ref{di}) in the continuum approximation
coincides with the sine-Gordon equation with losses \cite{landauer}.

\section{The continuum and the discrete JTL}
\label{dif}

Natural question is how good is
the continuum approximation used everywhere in this paper?
To answer this question let us  focus on Eq. (\ref{di}).
The continuum approximation   consists in promoting the discrete variable $n$ to the continuous variable $Z$ and approximating the discrete second order derivatives in the r.h.s. of (\ref{di}) by the continuous derivatives:
\begin{subequations}
\begin{alignat}{4}
\sin\varphi_{n+1}-2\sin\varphi_n+\sin\varphi_{n+1}
&=\frac{\partial^2 \sin\varphi}{\partial Z^2} \label{qq}\\
\varphi_{n+1}-2\varphi_n+\varphi_{n+1}
&=\frac{\partial^2\varphi}{\partial Z^2}.   \label{qq2}
\end{alignat}
\end{subequations}

To find the limits of  applicability of this approximation, let us
go one step further and consider the quasi-continuum  approximation
 \cite{kogan2}
 \begin{eqnarray}
\label{cqq2}
\sin\varphi_{n+1}-2\sin\varphi_n&+&\sin\varphi_{n+1}
=\frac{\partial^2\sin \varphi}{\partial Z^2}+\frac{1}{12}\frac{\sin\partial^4 \varphi}{\partial Z^4}.\nonumber\\
\end{eqnarray}
In this approximation,
the equation describing the localized travelling wave  is
\begin{eqnarray}
\label{system29}
\frac{1}{12\overline{U}^2}\frac{d^2\sin\varphi}{d \tau^2}
+\frac{C_J}{C}\frac{d^2\varphi}{d \tau^2 }+\frac{Z_J}{R_J} \frac{d\varphi}{d \tau }
=\overline{U}^2\varphi-\sin\varphi+F\nonumber\\
\end{eqnarray}
(compare with (\ref{system26})). So the  continuum approximation is applicable if either $C_J/C\gg 1$ or $Z_J/R_J\gg 1$.

Lossless JTL clearly corresponds to $R_J=\infty$. In addition,
while talking about  the lossless system in the present paper, we had in mind  Eq. (\ref{system29}) containing only the second term   in the l.h.s..
Previously \cite{kogan2}, we considered  unshunted JTL in the quasi-continuum
approximation, that is   Eq. (\ref{system29}) with only the first term in the l.h.s..
However, because of the similarity of the terms,  the kinks and the solitons obtained in the framework of these two considerations are qualitatively very similar.
(The quantitative differences  do exist. Thus the equation for the "soliton" velocity (\ref{solist})
is different from the equation for the soliton velocity (36) from Ref. \cite{kogan2}).

\section{Numerical integration of Eq. (\ref{system26})}
\label{nomer}

To integrate numerically Eq. (\ref{system26}) (for the sake of definiteness we consider the case $\varphi_1>0$), we should present it as an equation with the initial conditions. To do it let us first concentrate on the vicinity of the point $\varphi_1$. There the equation can be linearised and presented as
\begin{eqnarray}
\label{system33}
\frac{d^2\varphi}{d \tilde{\tau}^2 }
+\gamma\frac{d\varphi}{d\tilde{\tau}}
+\left(\cos\varphi_1-\overline{U}_{\varphi_2,\varphi_1}^2\right)
(\varphi-\varphi_1)=0.\nonumber\\
\end{eqnarray}
Solving Eq. (\ref{system33}) while taking into account first of the boundary conditions (\ref{12}) we obtain
\begin{eqnarray}
\label{333}
\varphi(\tilde{\tau})=\varphi_1-Ae^{\kappa\tilde{\tau}},
\end{eqnarray}
where
\begin{eqnarray}
\kappa=\sqrt{\gamma^2/4-\cos\varphi_1+\overline{U}_{\varphi_2,\varphi_1}^2}
-\gamma/2
\end{eqnarray}
and $A$ is an arbitrary positive constant.

Let us put in (\ref{333}) $\tilde{\tau}=0$. We obtain
\begin{eqnarray}
\label{333a}
\varphi(0)=\varphi_1-A
\end{eqnarray}
(for (\ref{333a}) to make sense, the constant $A$ should satisfy the condition $A\ll\varphi_2$). Differentiating (\ref{333}) with respect to $\tilde{\tau}$ and again putting $\tilde{\tau}=0$ we obtain
\begin{eqnarray}
\label{333b}
\left.\frac{d\varphi}{d\tilde{\tau}}\right|_{\tau=0}=-A\kappa.
\end{eqnarray}
Equations (\ref{333a}) and (\ref{333b}) we use as the initial conditions while numerically solving (\ref{system26}) for $\tau>0$.

\section{The elementary  particular solutions of the generalized dHD equation}
\label{ana}

 We consider the generalized dHD equation
\begin{eqnarray}
\label{44}
x_{\tau\tau}+\gamma x_{\tau}
=\alpha x\left(x^n-x_1\right)\left(x^n+x_3\right),
\end{eqnarray}
where $n$ is an integer, $\gamma$, $\alpha$,  $x_1,x_3$  are  positive constants.

Looking for an elementary particular solution of (\ref{44}) we try first to integrate the equation by quadrature. To do it let us  get read of the first derivative with respect to $\tau$ in (\ref{44}) by introducing new dependent variable
\begin{eqnarray}
\label{5.10}
x(\tau)=e^{-m\tau}w(\tau),
\end{eqnarray}
where  the  parameter $m$ will be determined later.
This change of variable turns  Eq. (\ref{44}) into
\begin{eqnarray}
\label{5.12}
w_{\tau\tau}
+(\gamma-2m)w_{\tau}=\alpha e^{-2nm\tau}w^{2n+1}\nonumber\\
-\alpha(x_1-x_3)e^{-nm\tau}w^{n+1}-\left(m^2-m\gamma+\alpha x_1x_3\right)w.
\end{eqnarray}
We see that the choice $m=\gamma/2$   would cancel the first derivative term in the equation, but the price is too high - the coefficients before the nonlinear terms in the equation would become explicitly $\tau$-dependent.

However we can   achieve our aim  in a different way.  Let us first kill the last term in   the r.h.s. of  (\ref{5.12}) by choosing
$m$ satisfying the equation
\begin{eqnarray}
\label{mumu}
m^2-m\gamma+\alpha x_1x_3=0.
\end{eqnarray}
Let us also make the change of the independent variable
\begin{eqnarray}
\label{5.13}
\tau\to z(\tau)=e^{-nm\tau}.
\end{eqnarray}
After that,  Eq. (\ref{5.12}) becomes
\begin{eqnarray}
\label{5.15}
n^2m^2w_{zz}+n\left[(n+1)m^2-\alpha x_1x_3\right]\frac{w_z}{z}\nonumber\\
=\alpha  w^{2n+1}-\alpha (x_1-x_3) \frac{w^{n+1}}{z}.
\end{eqnarray}

\subsection{Integration  by quadrature}

Consider the particular case $x_3=x_1$.
In this case   (\ref{5.15})  can be easily integrated by quadrature, provided
the coefficient before the first derivative is equal to zero,
i.e. $m$ is
\begin{eqnarray}
\label{meme}
m=\sqrt{\frac{\alpha}{n+1} }x_1.
\end{eqnarray}
Note that
the condition of compatibility of (\ref{meme}) and (\ref{mumu}) (the latter with $x_3=x_1$) is
\begin{eqnarray}
\label{cond}
\gamma=(n+2)m.
\end{eqnarray}
The  condition being assumed,  Eq.(\ref{5.15}) becomes
\begin{eqnarray}
\label{5.19}
w_{zz}-\frac{n+1}{n^2x_1^2}w^{2n+1}=0
\end{eqnarray}
(compare  (\ref{5.19}) with (\ref{44})).
Multiplying Eq.(\ref{5.19}) by $w_{z}$ and integrating with respect to $z$ we have
\begin{eqnarray}
\label{5.20}
w_{z}^2-\frac{1}{n^2x_1^2}w^{2n+2}=2E,
\end{eqnarray}
where $E$ is the integration  constant.

\subsection{The elementary solutions}
\label{old}

Let us demand that the solution of  (\ref{44}) satisfies the   boundary conditions
\begin{eqnarray}
\label{condi}
\lim_{\tau\to-\infty}x(\tau)=x_1^{1/n},\hskip 1 cm \lim_{\tau\to+\infty}x(\tau)=0.
\end{eqnarray}
Note that if  the solution of (\ref{44}) $x(\tau)$ should exist and remain finite for $\tau\in(-\infty,+\infty)$,  Eq. (\ref{condi}) are the only possible  boundary conditions  at $\tau=\pm\infty$ (apart from changing $x_1$ to $x_3$).  The trajectory should start in the infinite past in the unstable fixed point  and end  in the infinite future in the stable fixed point.

From (\ref{condi}) follows that the solution of (\ref{5.20})
should satisfy the boundary condition
\begin{eqnarray}
\label{conditor}
\lim_{z\to \infty}z^{1/n}w(z)=x_1,
\end{eqnarray}
and we should substitute $E=0$ into (\ref{5.20}). This gives us the opportunity to integrate  the equation not by quadrature, but in terms of elementary function:
\begin{eqnarray}
\label{5.22}
w(z)=\frac{x_1^{1/n}}{\left(1+z\right)^{1/n}}.
\end{eqnarray}
Finally substituting (\ref{5.22}) into  (\ref{5.10}), we obtain  elementary  particular solution of (\ref{44})
\begin{eqnarray}
\label{fii2}
x(\tau)=\frac{x_1^{1/n}.}
{\left[\exp\left(nm\tau\right)+1\right]^{1/n}},
\end{eqnarray}
which exists, provided the constants of  (\ref{44})
are connected by the relation (\ref{cond}).

Now we have a pleasant surprise: the elementary solution (\ref{5.22}) solves
(\ref{5.15}) also when $x_3\neq  x_1$. In fact,
substituting (\ref{5.22}) into (\ref{5.15}) we obtain
\begin{eqnarray}
\label{44b}
\frac{(n+1)m^2}{(1+z)^2}- \frac{(n+1)m^2-\alpha x_1x_3}
{z(1+z)}\nonumber\\
= \frac{\alpha x_1^2}{(1+z)^2}-  \frac{\alpha (x_1-x_3)x_1}{z(1+z)},
\end{eqnarray}
and mysteriously
the terms in the l.h.s and in the r.h.s. of the equation are equal pairwise, provided $m$ is given by (\ref{meme}). Hence (\ref{fii2}) is
the elementary solution of
(\ref{44}) valid  also when $x_3\neq  x_1$).
The condition of compatibility of (\ref{mumu}) and (\ref{meme}) in the general case is
\begin{eqnarray}
\label{g2}
\gamma=\sqrt{\frac{\alpha}{n+1}}[x_1+(n+1)x_3]
\end{eqnarray}
 (compare with(\ref{cond})).

\subsection{The strongly asymmetric case}

For $x_3 \gg  x_1$ there exists additional elementary solution of (\ref{44}). In this case the equation can be approximated as
 \begin{eqnarray}
\label{44c}
x_{\tau\tau}+\gamma x_{\tau}=\alpha x_3 x\left(x^n-x_1\right)\nonumber\\
=\alpha x_3 x\left(x^{n/2}-x_1^{1/2}\right)\left(x^{n/2}+x_1^{1/2}\right).
\end{eqnarray}
Comparing (\ref{44c}) with (\ref{44}) (and keeping in mind the elementary solution of the  latter (\ref{fii2}), we obtain   the elementary  solution of (\ref{44c})
\begin{eqnarray}
\label{fon}
x(\tau)=\frac{x_1^{1/n}}
{\left[\exp\left(nm'\tau\right)+1\right]^{2/n}}.
\end{eqnarray}
where
\begin{eqnarray}
\label{g22}
m'=\sqrt{\frac{\alpha x_3x_1}{2(n+2)}},\hskip 1cm \gamma=(n+4)m'.
\end{eqnarray}

\subsection{Abel equation}

The  obtained elementary solutions  become more transparent    after we notice that  introducing new dependent variable $p$
\begin{eqnarray}
\label{px}
p=x_{\tau}
\end{eqnarray}
 and considering $x$ as the independent variable, we may write down (\ref{44})  as Abel equation of the second kind   \cite{polyanin}
\begin{eqnarray}
\label{ab}
pp_{x}+\gamma p=\alpha x\left(x^n-x_1\right)\left(x^n+x_3\right).
\end{eqnarray}
 For $\gamma$ and $\alpha$ connected by the formula (\ref{g2}), Eq. (\ref{ab})  has  the  particular solution
\begin{eqnarray}
\label{ab2}
p= \sqrt{\frac{\alpha}{n+1}}x\left(x^n-x_1\right)
\end{eqnarray}
Substituting $p(x)$ into (\ref{px}) and integrating thus obtained differential equation we obtain (\ref{fii2}).

 Approximating (\ref{ab}) for $x_3\gg x_1$ as
\begin{eqnarray}
\label{ab5}
pp_{x}+\gamma p&=&\alpha x_3x\left(x^n-x_1\right)\\
&=&\alpha x_3x\left(x^{n/2}-x_1^{1/2}\right)\left(x^{n/2}+x_1^{1/2}\right),\nonumber
\end{eqnarray}
we obtain  from (\ref{ab2}) the particular solution of (\ref{ab5})
\begin{eqnarray}
\label{ab22}
p= \sqrt{\frac{2\alpha x_3}{n+2}}x\left(x^{n/2}-x_1^{1/2}\right),
\end{eqnarray}
and from (\ref{g2})  the connection between $\gamma$ and $\alpha$ in the present case; the latter turns out to be identical to that given by  (\ref{g22}).
Substituting  obtained $p(x)$ into (\ref{px}) and integrating thus obtained differential equation we obtain  (\ref{fon}).

\end{appendix}

\begin{acknowledgments}

I am grateful to  A. Abanov, M. Goldstein, B. Malomed and
J. Cuevas-Maraver for the discussion. I am also very grateful to Donostia International Physics Center (DIPC)
for the hospitality during my visit, when the paper was finalised.

\end{acknowledgments}
%\section{Conflict of Interest}

%The author declare no conflict of interest.

\end{document}